# Variable stars in the field of the old open cluster NGC 2243 [1]

Janusz Kaluzny

Warsaw University Observatory

jka@sirius.astrouw.edu.pl

Wojciech Krzemiński

Carnegie Observatories, Las Campanas Observatory

wojtek@roses.ctio.noao.edu

Beata Mazur

Copernicus Astronomical Center

batka@camk.edu.pl

## Abstract

We report results of photometric monitoring of the NGC 2243 field in $V$ and $I$ filters. Two W UMa-type systems and three detached eclipsing binaries were discovered. Both contact systems are likely members of the cluster. One of detached binaries – variable V1 – has a period P=1.188 day. This variable is located at the cluster's turnoff in the color-magnitude diagram. Determination of parameters for components of V1 can provide direct information about properties of turnoff stars in NGC 2243 as well as to allow direct determination of cluster's distance. A background RR Lyr variable was found in the monitored field. The reddening of NGC 2243 is estimated at $E(V - I) = 0.10 \pm 0.04$ based on the $V - I$ color exhibited by this star at minimum light. The observed distribution of colors for the background halo stars shows a cutoff at $(V - I) \approx 0.60$ what implies $E(V - I) \leq 0.12$ for the cluster.

## 1.  Introduction

This paper is a contribution to the on-going systematic search for short-period variables in open and globular clusters which has been started by our team in December 1989 and is based on time resolved CCD photometry. The main goals of this program were described by Kaluzny *et al.* (1993). References to results published before 1995 can be found in a paper on NGC 7789 (Jahn *et al.* 1995). Recently we published a paper about variables in Cr 261 (Mazur *et al.* 1995)

NGC 2243 is an old open cluster located at relatively high galactic latitude ($b = -18°$). Its pioneering study, based on the photographic photometry, was conducted by Hawarden (1975). The color-magnitude diagram of the cluster shows a clearly marked sequence of binary stars (Bonifazi *et al.* 1990, Bergbush *et al.* 1991). These properties make NGC 2243 an attractive target for eclipsing binaries search.

---

[1] Based on observations collected at Las Campanas Observatory

## 2. Observations and data reduction

The field of NGC 2243 was surveyed for variable stars with the 1-m Swope telescope at Las Campanas Observatory. The observational data were collected during five runs conducted between December 1990 and December 1993. Four different CCD cameras were used:

$1024 \times 1024$ Tektronics CCD with scale of 0.61 $arcsec/pixel$ – TEK2 camera
$1024 \times 1024$ Tektronics CCD with scale of 0.69 $arcsec/pixel$ – TEK1 camera
$2048 \times 2048$ Tektronics CCD with scale of 0.61 $arcsec/pixel$ – TEK3 camera
$2048 \times 2048$ Ford Aerospace CCD with scale of 0.435 $arcsec/pixel$ – FORD camera.

No log of observations is presented here. However, time distribution of our observations can be inferred from the light curves of variables, which were submitted in tabular form to the editors of A&A (see appendix A). We obtained a total of 367 frames in the $V$-band and 197 in the $I$-band. Exposure times ranged from 3 to 7 minutes with 4 minutes being the most frequent value. Most of data were collected on 7 nights between 1992 February 3 and 15 (UT).

The preliminary processing of the raw data was made with IRAF[2]. Stellar photometry was extracted using DAOPHOT/ALLSTAR (Stetson 1987, 1991) and DoPHOT (Schechter *et al.* 1993) programs. All sets of images suffered from some positional dependence of the point spread function (PSF). DAOPHOT/ALLSTAR was used to analyze images collected with TEK1 or TEK2 cameras. A PSF varying linearly across an image was adopted. To improve reliability of photometry we adopted procedure which is described in Kaluzny *et al.* (1995a). Images obtained with TEK3 and FORD cameras were analyzed with DoPHOT. To cope with effects of variable PSF we used procedure described in Kaluzny et al. (1995b). A separate data base was prepared for each combination of CCD camera and filter. Photometry for all 8 data bases was transformed to the standard $VI$ system. First each data set was transformed from the instrumental to the standard system using transformation derived for a given chip. Subsequently small shifts of the zero points were applied to tie all photometry to the system defined by data set obtained with the TEK2 camera (see section 6 for more details).

## 3. The search for variables

All data bases were independently searched for variable stars. Two methods of selecting potential variables were used. First we picked up stars with light curves showing large dispersion as compared with other stars of similar magnitude. To illustrate quality of our photometry we present in Fig. 1 a plot $rms$ versus the average magnitude for light curves from the data base corresponding to filter $V$ and TEK2 camera. The second method used was scanning of light

---

[2]IRAF is distributed by the National Optical Astronomy Observatories, which are operated by the Associations of Universities for Research in Astronomy, Inc., under cooperative agreement with the National Science Foundation

curves with a filter designed to detect eclipse-like events. Light curves of all candidate variables were subject to detailed examination. After elimination of spurious variables we were left with 6 clearly variable stars. Table 1 lists the equatorial and rectangular coordinates of these stars. The rectangular coordinates correspond to the $V$-band image obtained with a TEK3 camera. This image is available from the data base maintained by A&A (see Appendix A). A transformation from rectangular to the equatorial system was derived based on positions of 12 stars from the GS Catalogue (Lasker *et al.* 1988). The errors of equatorial coordinates listed in Table 1 are about 1 arcsec.

## 4. Light curves and period determination

To determine periods of newly discovered variables we used the analysis of variance method, as described by Schwarzenberg-Czerny (1989, 1991). It was possible to determine periods for 4 out of 6 variables: two W UMa-type systems, one eclipsing binary with EA-type light curve and one RR Lyr star. Phased light curves of these stars are shown in Fig. 2. The light curve of V1 shows two very similar minima when phased with P=1.19 $d$. Adoption of P=0.595 $d$ for this variable leads to light curve with an unacceptably wide primary minimum lasting $\Delta\phi \approx 0.25$ and no trace of the secondary eclipse. Hence, period P=0.595 $d$ can be excluded for V1. Variable V3 showed an unstable light curve with large seasonal changes. Most of scatter visible in its light curve is due to secular changes of its brightness at a given phase. Such behaviour is in fact quite common among W UMa-type binaries. The best known example of a contact system with very unstable light curve is TZ Boo (Hoffmann 1980).

Two eclipsing episodes were observed for variable V4. This star is located in the outskirts of the cluster. It was not observed on nights when we used a TEK2 camera. In Fig. 3 we show the light curve of V4 obtained in February 1992. If two observed eclipses of V4 correspond to the same minimum then its orbital period is either P=2.84 $d$ or P=1.42 $d$. Periods shorter than 1.42 days can be ruled out. In addition to two eclipses observed in February 1992 we observed also an ending part of an eclipse of V4 on 1993 Dec 21 (UT).

Two likely eclipsing episodes were observed for variable V5. On 1992 Feb 11 (UT) a descending branch of eclipse was caught while a small fragment of a central part of an eclipse was observed on 1990 Dec 27 (UT). A relevant fragments of the light curve are shown in Fig. 4. Our photometry revealed moreover some secular changes of the out-of-eclipse brightness of V5. In 1990, 1991 and 1992 seasons the variable showed $V_{max} \approx 16.30$. On six nights in December 1993 its brightness was $V_{max} \approx 16.20$.

Some basic characteristics of six newly discovered variables are given in Table 2.

## 5. Control fields photometry

Table 1: Rectangular and equatorial coordinates of variables identified in the field of NGC 2243. Rectangular coordinates can be used to identify all objects on the image included in the set of supplementary data accompanying this paper (see Appendix A).

| No | $X$ | $Y$ | RA(2000) | DEC(2000) |
|----|-----|-----|----------|-----------|
| V1 | 917.2 | 968.95 | 6:29:35.47 | -31:16:53.0 |
| V2 | 881.9 | 984.9 | 6:29:33.80 | -31:17:02.8 |
| V3 | 1115.3 | 1110.6 | 6:29:44.87 | -31:18:18.8 |
| V4 | 371.7 | 836.5 | 6:29:09.62 | -31:15:33.0 |
| V5 | 1349.2 | 1308.2 | 6:29:55.98 | -31:20:18.4 |
| V6 | 876.6 | 1075.1 | 6:29:33.56 | -31:17:57.7 |

Table 2: Parameters of variables discovered in the field of NGC 2243. $V_{max}$ and $(V - I)$ are magnitude and color at the maximum light. For V5 we list parameters corresponding to the 1991-1992 period. The last two columns give adopted ephemerides. The period is given in days. $T_0$ gives the time of minimum light for eclipsing binaries and the time of maximum light for the RR Lyr variable.

| ID | Type | $V_{max}$ | $V - I$ | P | $T_0$ JDhel 2450000+ |
|----|------|-----------|---------|---|----------------------|
| V1 | EA | 16.33 | 0.59 | 1.188506 | 8550.7996 |
| V2 | EW | 17.82 | 0.84 | 0.285300 | 8224.8510 |
| V3 | EW | 16.66 | 0.61 | 0.356455 | 8224.6671 |
| V4 | EA | 14.02 | 0.54 | | |
| V5 | EA | 16.28 | 0.65 | | |
| V6 | RR | 16.87 | 0.29 | 0.586591 | 8224.4532 |

On nights of 1993 December 28-31 we monitored two "control" fields offsetted by $\Delta l = \pm 0.5$ deg from the center of NGC 2243, respectively ($l$ denotes galactic longitude). Each of these fields was monitored for a total of about 5 hours. We used the TEK2 camera and most of images were taken in the $I$ band. Some thin cirrus was present on a sky during these observations but good seeing allowed to obtain relatively deep photometry. The data were processed following procedures applied to images of NGC 2243. A search for variable stars gave a negative result for both control fields. Our data indicate that control fields do not contain any W UMa-type systems brighter than $I \approx 19$ mag and total amplitude of light variations exceeding 0.1 mag. The short time-base of our data does not allow to reach any firm conclusions about presence or absence of other types of variables in the control fields.

## 6. The color-magnitude diagram

Photometry extracted from images obtained with the TEK2 camera was averaged to produce a color-magnitude diagram (CMD) for the monitored field. Photometry based on relatively poor images was not taken into account. We used data extracted from 74 images in the $V$ band and from 148 images in the $I$ band. For each frame stars showing relatively large errors of photometry for their magnitude were flagged. Such poor measurements were rejected while calculating the average magnitudes. Only stars with at least 9 acceptable measurements in the $V$ band and at least 17 acceptable measurements in the $I$ band were retained. The final lists with averaged photometry contained 1728 and 2591 stars for the $V$ and $I$ band, respectively. Of these stars 1697 objects were present in both lists. Transformation from the instrumental to the standard system was derived based on observations of several Landolt (1992) fields. The systematic uncertainty in the adopted calibration is about 0.020 mag for $V$ and about 0.015 for $V - I$. The color-magnitude diagram (CMD) based on the data collected with a TEK2 camera is shown in Fig. 5. We emphasise that completeness of the photometry was sacrificed for the sake of its accuracy. Hence, some stars, especially ones from the central region of the cluster, were not included in Fig. 5. The presented CMD shows the same features which have already been discussed extensively by Bonifazi *at al.* (1990) and Bergbush *et al.* (1991). A group of candidate blue stragglers and a sequence of binary stars located above the cluster main-sequence can be noted. Our photometry covers larger field and seems to be more accurate and deeper in comparison with previous studies of NGC 2243. In Fig. 6 we show CMD's for two groups of stars: those lying within radius $R = 164$ arcsec from the cluster center, and those lying in an outer part of the observed field at $R > 301$ arcsec. The inner circle and the outer region cover the same areas on the sky. In Fig. 7 we plot the projected density distribution for stars with reliable photometry and brighter than $I = 20.4$. Some interesting conclusions can be reached on examination of Figs. 5-7:

- The cluster main sequence can be traced down to $V \approx 21.8$, which is the limiting magnitude of our photometry. Also the sequence of binaries extends down to the faintest observed magnitudes.

- The density profile shown in Fig. 7 flattens beyond $R \approx 4.0$ arcmin. However, the cluster main sequence can be easily distinguished in the CMD for the outer field (Fig. 6b). Hence the cluster radius is well above 4 arcmin. In fact van den Bergh (1977) was able to trace the cluster out to a radius of 6.3 arcmin.

- The CMD for the outer field (Fig. 6b) contains no stars with $V < 15.5$ and $V - I > 0.7$. Hence, most of objects forming scattered subgiant/giant branch in a CMD for the inner field (Fig. 6a) are cluster members. The large scatter observed on the subgiant branch of NGC 2243 can be attributed to binarity of a significant fraction of its stars.

- There is a sizeable population of field stars visible below the cluster main sequence (see Fig. 5). Their color shows a blue cutoff at $V - I \approx 0.60$. This cutoff corresponds to the turnoff of stars from the halo population and its location can potentially provide an upper limit on the cluster reddening. Just to illustrate the proposed method we used $VI$ photometry of M68 published by Walker (1994). M68 is one of the most metal poor galactic globular clusters and its turnoff is located at $(V - I)_0 \approx 0.48$. It may be assumed that halo stars observed in the background of NGC 2243 are not older and have metallicity not lower than stars in M68. Consequently we obtain $E(V - I) = 0.12$ as an upper limit on the reddening of NGC 2243.

## 7.  Cluster membership of the variables

Figure 8 shows location of 6 newly identified variables on the cluster CMD. For all variables but V6 the marked positions correspond to magnitude and color at the maximum light. The position marked for RR Lyr star V6 corresponds to its intensity-averaged magnitude and average color. We obtained for this star $< V > = 17.456$ and $< I > = 16.952$. Variable V6 is clearly a background object located well behind the cluster. It is known that the mean minimum-light $(V - I)$ color of RRab variables is $0.58 \pm 0.03$ with little or no dependence on metallicity (Mateo et al. 1995). For V6 we obtained $(V - I)_{\min} = 0.68 \pm 0.01$ what implies $E(V - I) = 0.10 \pm 0.04$ and $E(B - V) = 0.08 \pm 0.04$. This sets formally an upper limit on the reddening of NGC 2243. This result is in good agreement with Hawarden (1975) and Bonifazi et al. (1990) who obtained $E(B - V) = 0.06$ and $E(B - V) = 0.06 - 0.08$, respectively.

Contact binaries V2 and V3 are located slightly above the main sequence in the cluster CMD. We have applied the absolute brightness calibration established recently by Rucinski (1994) to estimate their absolute magnitudes. Rucinski's calibration gives $M_V$ as a function of period and color of a contact system. We obtained $M_V^{cal} = 4.79$ and $M_V^{cal} = 3.53$ for V2 and V3, respectively. The estimates of the apparent distance modulus of NGC 2243 range from 12.7 to 13.05 and we may adopt here $(m - M)_V = 12.90 \pm 0.20$. Assuming cluster membership we obtain $M_V^{obs} = 4.92 \pm 0.20$ and $M_V^{obs} = 3.76 \pm 0.20$ for V2 and V3, respectively. Good agreement between values of $M_V^{cal}$ and $M_V^{obs}$ supports hypothesis that V2 and V3 are located at the same distance as NGC 2243.

Eclipsing binary V4 is located almost 2 magnitudes above the top of main sequence of NGC 2243. It is located far away from the cluster center at a projected distance of $R = 4.9$ arcmin. This variable is probably just a foreground object not belonging to the cluster. Binaries V1 and V5 are likely cluster members as they occupy positions among photometric binaries in the cluster CMD. Moreover, they are located in the central part of NGC 2243.

## 8.   Summary

The most interesting result of our survey for variable stars in NGC 2243 is discovery of the detached eclipsing binary V1 which is located near the cluster turnoff in the color-magnitude diagram. The equal depths of eclipses for stars in this region of the CMD implies equal effective temperatures of components and, most likely, similarity of the other parameters, such as radii, masses, etc.. The light curve is flat outside eclipses what indicates that system is well detached. By combining precise radial velocity curves with accurate photometry one would be able to determine absolute parameters for both components of V1. That in turn would allow to estimate age and distance to NGC 2243 using procedures which are to large extent free from uncertainties involved in a standard isochrone fitting methods. The masses of spectroscopic binaries are proportional to the cube of the radial velocity amplitudes. Hence to obtain astrophysically interesting results one needs to measure the radial velocity amplitudes $K_1$ and $K_2$ with 1-2% accuracy. Such an accuracy is attainable with the currently available instrumentation provided that one has access to the 4-m class telescopes.

We identified two contact binaries which are likely members of the cluster. The relative frequency of W UMa stars in NGC 2243 is significantly lower than the same frequency for NGC 188. The latter cluster is known to harbor at least 7 contact binaries. Let us denote by $N_{6.5}$ the number of stars with $M_V < 6.5$ and by $N_{EW}$ the number of contact binaries. For NGC 188 we have $N_{6.5}/N_{EW} \approx 43$ while the same quantity for NGC 2243 is about 500. For NGC 2243 we limited our attention to the field monitored with TEK2. No corrections for the field stars contamination were taken into account but this has little effect on the derived values of $N_{6.5}/N_{EW}$. In fact, the relative frequency of contact binaries among NGC 2243 F-K dwarfs does not differ from the frequency observed for the field stars.

## 9.   Acknowledgements

JK was supported by the Polish KBN grant 2P03D008-08 and by the NSF grants AST-9313620 and AST-9216494 to Bohdan Paczyński. BM was supported by the Polish KBN grant 2-P30401307. We thank Slavek Rucinski for useful comments about early version of this paper

## 10.   Appendix A

Tables containing the light curves of all newly discovered variables as well as $VI$ photometry presented in Fig. 5 are published by A&A at the Centre de Donnés de Strassbourg, where they are available in electronic form: See the Editorial in A&A 1993, Vol. 280, page E1. We have also submitted an image of NGC 2243 field to the data base which should allow identification of all the new variables.

## 11.    Appendix B

A single pair of $V$ and $I$ frames obtained with the TEK3 camera was used to produce the CMD for the field of NGC2243. This CMD is presented in Fig. 9. Derived photometry is less accurate than that presented in Fig. 5 but it covers significantly larger field. The data presented in Fig. 9 were submitted to A&A (see Appendix A).

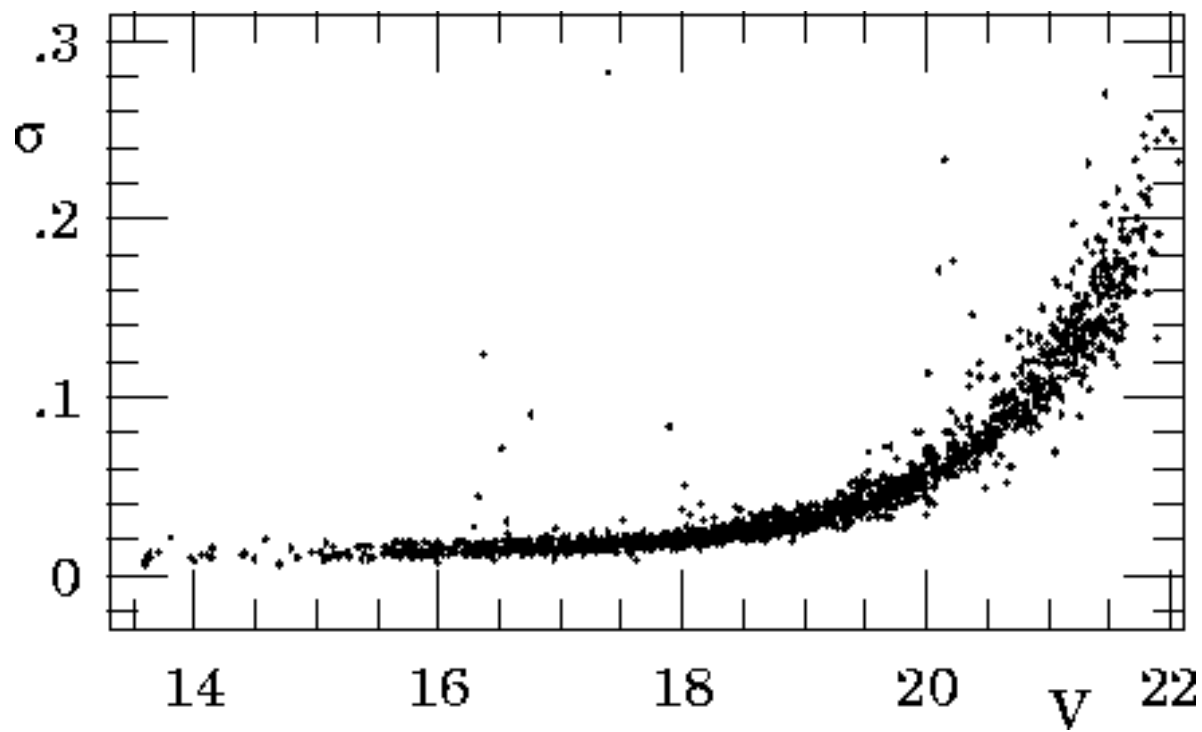

Fig. 1.— Standard deviations of the light curves versus the average $V$ magnitude for stars from the data base for filter $V$ and TEK2 camera.

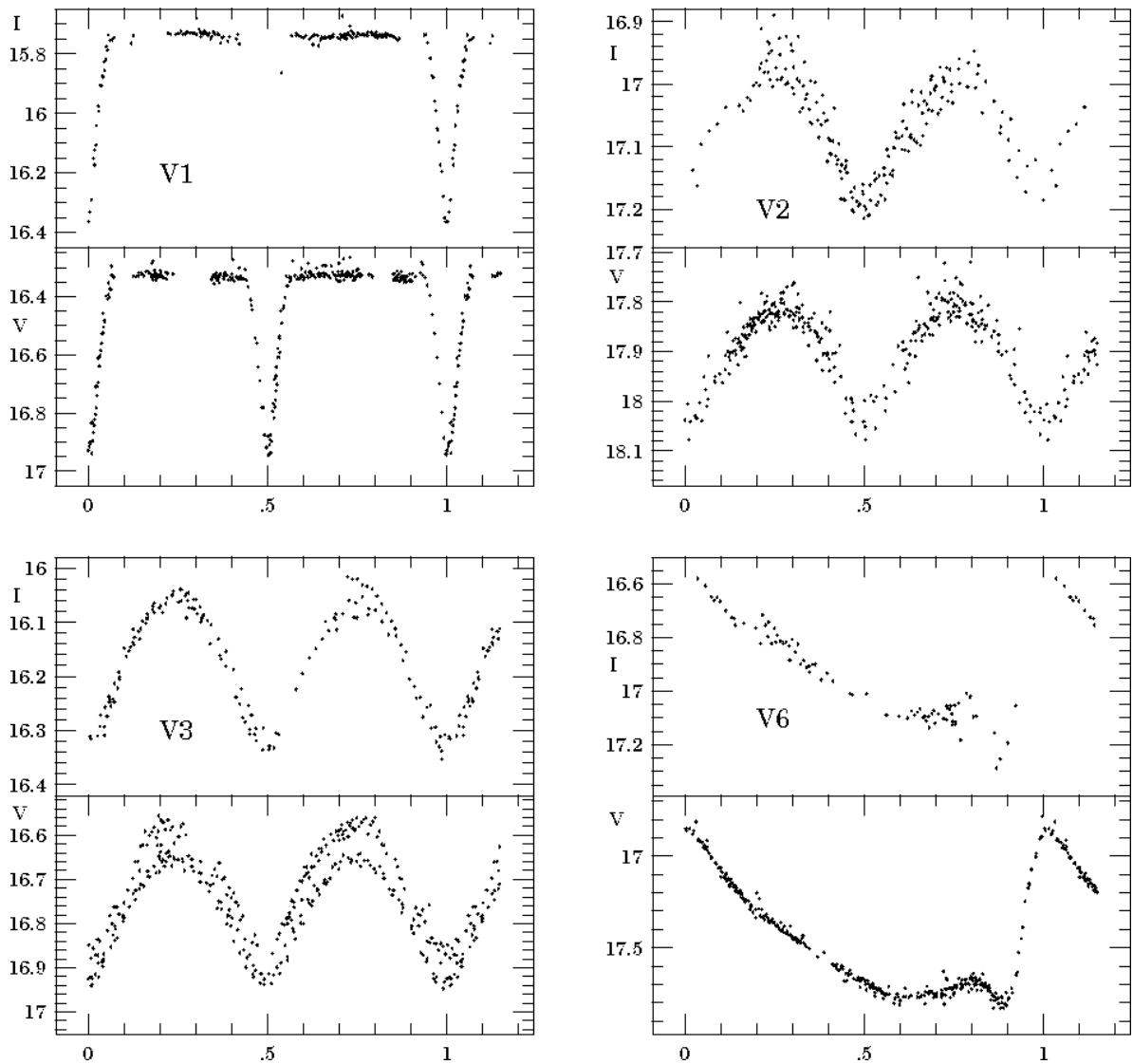

Fig. 2.— Phased light curves for variables V1-3 and V6 discovered in the field of NGC 2243. For each star the $I$-band (upper) and $V$-band (lower) light curves are shown.

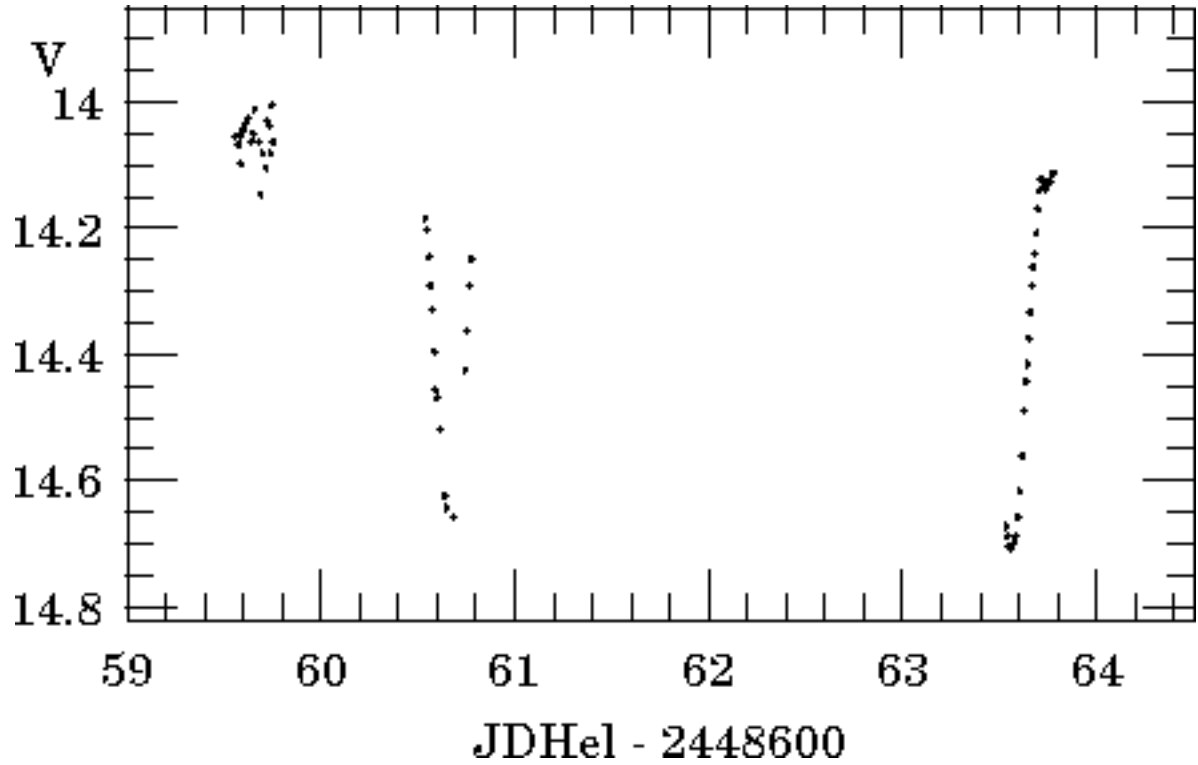

Fig. 3.— Sections of an unphased light curve of variable V4 discovered in the field of NGC 2243. Only data obtained on three nights in February 1992 are shown in this plot.

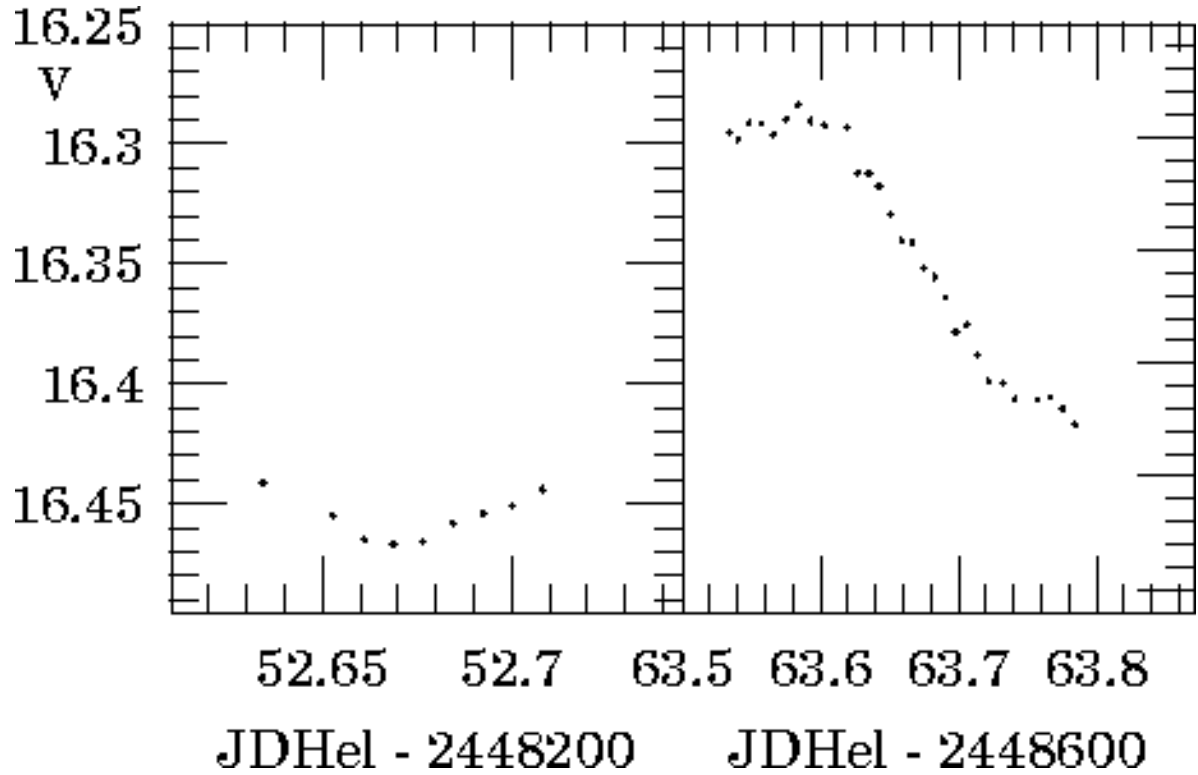

Fig. 4.— Sections of an unphased light curve of variable V5 discovered in the field of NGC 2243. Two likely eclipse events are shown.

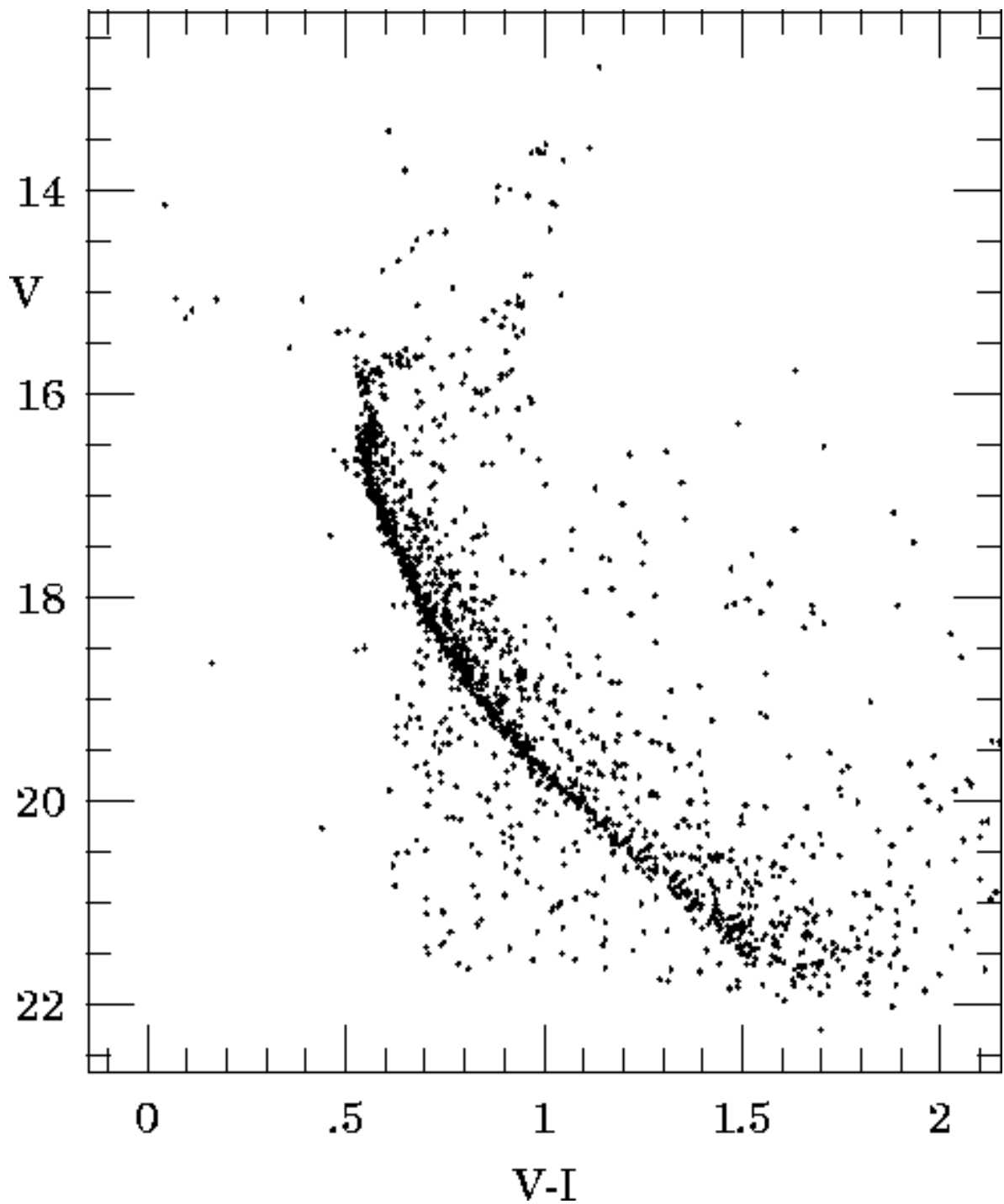

Fig. 5.— The CMD for the field of NGC 2243 monitored with the TEK2 camera.

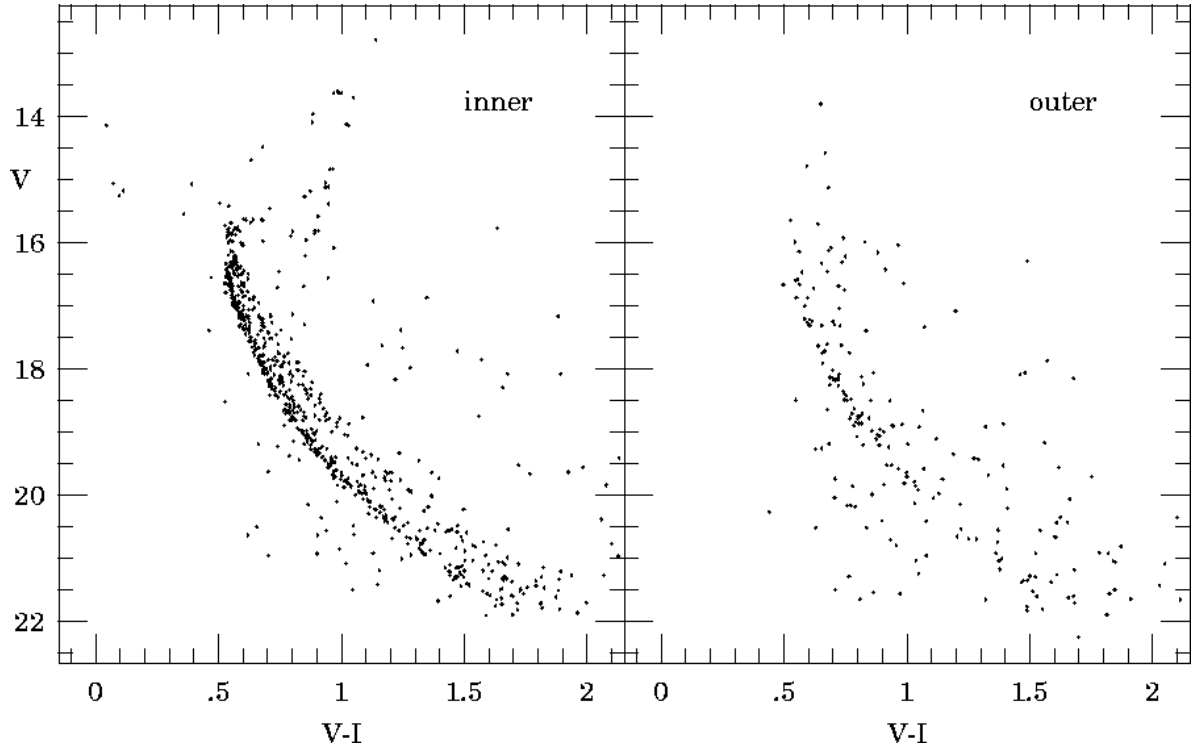

Fig. 6.— The CMD's for stars located at distances $R < 164$ arcsec (a: left panel) and $R > 301$ arcsec (b: right panel) from the projected center of NGC 2243. Note a clearly marked sequence of binaries visible on a left panel.

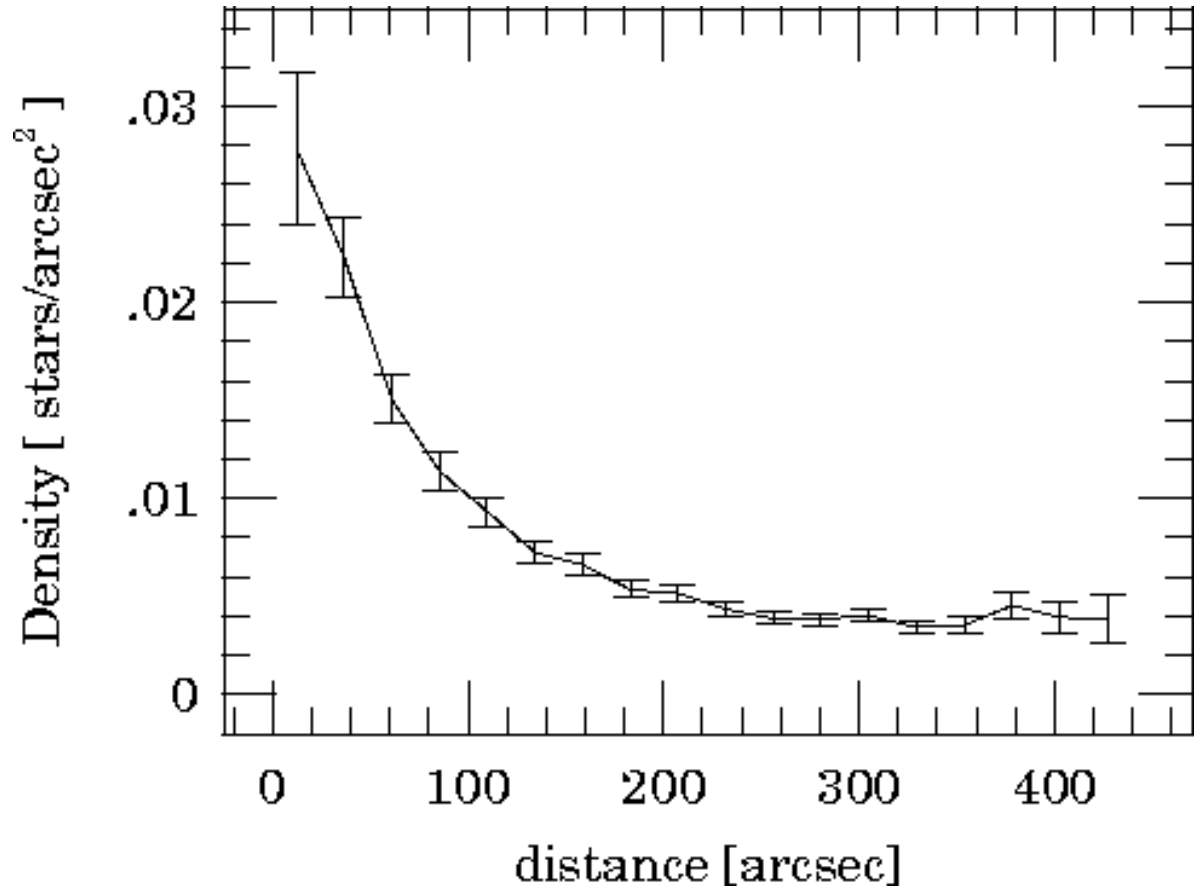

Fig. 7.— Projected star density as a function of radial distance from the center of NGC 2243 for objects with $I < 20.4$.

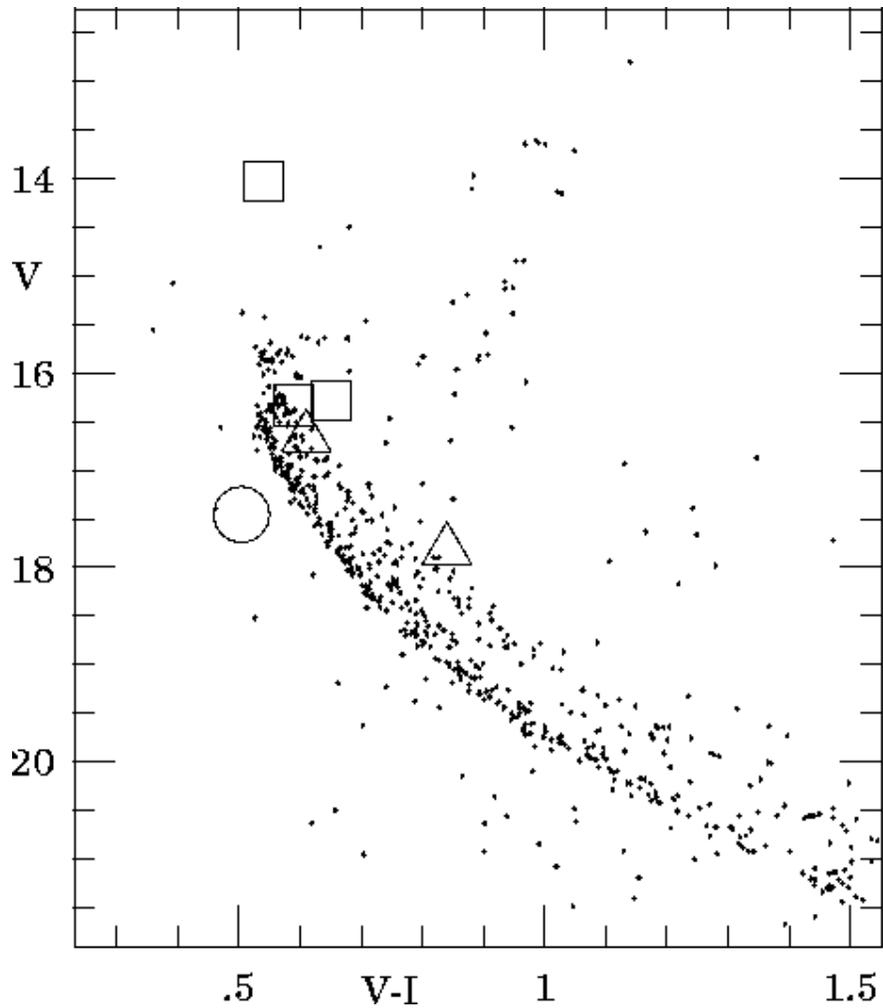

Fig. 8.— The CMD for the central part of NGC 2243 with the positions of the newly discovered variables marked. The triangles represent contact binaries, the squares EA-type eclipsing binaries, and the open circle an RR Lyr star. Position marked for RR Lyr star V6 corresponds to its average color and magnitude. For remaning variables we plotted $V_{max}$ and $(V - I)_{max}$. V6 is located well behind the cluster at galactocentric distance $r \approx 30.4$ kpc.

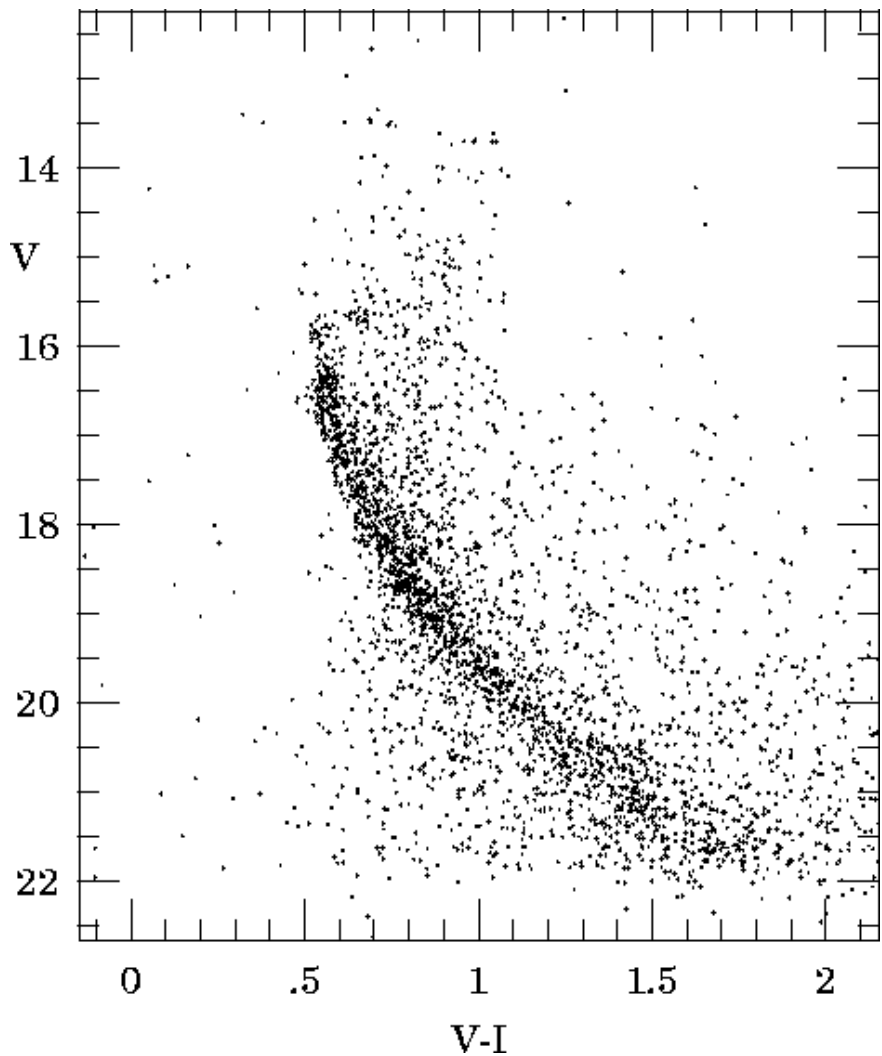

Fig. 9.— The CMD for the field of NGC 2243 monitored with the TEK3 camera.